\documentclass[11pt,a4paper]{article}
\usepackage[utf8]{inputenc}
\usepackage{amsmath}
\usepackage{amssymb}
\usepackage{cite}
\usepackage{graphicx}
\usepackage{bbm}
\usepackage{xcolor}
\newcommand{\be}{\begin{equation}}  
\newcommand{\ee}{\end{equation}}  
\newcommand{\ol}[1]{\overline{#1}}
\newcommand{\hc}{+\,\mathrm{h.c.}}
\newcommand{\vev}[1]{\langle #1 \rangle}

\newcommand{\SU}[1]{\ensuremath{\mathrm{SU}(#1)}}
\newcommand{\U}[1]{\ensuremath{\mathrm{U}(#1)}}
\newcommand{\into}{\ensuremath{\,\rightarrow\,}}

\newcommand{\tensor}{\ensuremath{\otimes}}

\newcommand{\chp}{\ensuremath{{ \psi^{+} } } }
\newcommand{\chpm}{\ensuremath{{ \psi^{\pm} } } }
\newcommand{\chpp}{\ensuremath{{ \psi^{++} } }}
\newcommand{\chm}{\ensuremath{{\psi^{-} } }}
\newcommand{\chmm}{\ensuremath{{ \psi^{--} } }}
\renewcommand{\d}{\ensuremath{\mathrm{d}}}

\newcommand{\mev}{\ensuremath{\mathrm{~MeV}}}
\newcommand{\gev}{\ensuremath{\mathrm{~GeV}}}
\newcommand{\mm}{\ensuremath{\mathrm{~mm}}}

\title{
\textbf{Next-to-minimal dark matter at the LHC} 
\vspace{2ex}
}

\author{\large A.~Bharucha$^a$, F.~Br\"ummer$^{b}$ and N.~Desai$^{b}$\\[1ex]
\textit{\normalsize $^a$Aix Marseille Univ, Universit\'e de Toulon, CNRS, CPT, Marseille, France}\\ 
\textit{\normalsize 13288 Marseille, France}\\
\textit{\normalsize $^b$ LUPM, UMR5299, Universit\'e de Montpellier and CNRS}\\
\textit{\normalsize 34095 Montpellier, France}\\[1ex]
\vspace{3ex}
}
\date{}
\begin{document}

\maketitle

\begin{abstract} \noindent
We examine the collider signatures of a WIMP dark matter scenario comprising a singlet fermion and an SU(2) $n$-plet fermion, with a focus on $n = 3$ and $n = 5$. The singlet and $n$-plet masses are of the order of the electroweak scale. The $n$-plet contains new charged particles which will be copiously pair-produced at the LHC. Small mixing angles and near-degenerate masses, both of which feature naturally in these models, give rise to long-lived particles and their characteristic collider signatures. In particular, the $n = 5$ model can be constrained by displaced lepton searches independently of the mixing angle, generically ruling out 5-plet masses below about 280 GeV. For small mixing angles, we show that there is a parameter range for which the model reproduces the observed thermal relic density but is severely constrained by disappearing track searches in both the $n = 3$ and the $n = 5$ cases. The $n = 3$ model is further constrained by soft di-lepton searches irrespective of whether any of the new particles are long-lived.
\end{abstract}

\section{Introduction}

Weakly interacting massive particles (WIMPs) continue to be promising dark matter candidates, even though the simplest WIMP models are increasingly constrained by direct and indirect detection experiments and by LHC search limits. Defining a WIMP in the narrower sense to be a particle with actual electroweak gauge interactions (as opposed to a generic particle with an electroweak-scale mass and e.g.~a Higgs portal coupling), the simplest WIMP dark matter model is arguably Minimal Dark Matter \cite{Cirelli:2005uq}: a single fermionic or scalar $n$-plet of $\SU{2}_L$, where $n$ is chosen such that its electrically neutral component is an effectively stable thermal relic. However, the favoured case of a fermionic $5$-plet \cite{Cirelli:2009uv} has a mass of the order of $10$ TeV, well beyond the reach of the LHC and probably of even a future 100 TeV collider \cite{Golling:2016gvc}. It might eventually be possible to probe this model with CTA \cite{Cirelli:2015bda}.

A next-to-minimal extension of the fermionic $n$-plet model can provide an effective description of dark matter and its coannihilation partners below the TeV scale\cite{Bharucha:2017ltz}. Dark matter is now mostly an electroweak singlet $\chi$ with a small admixture of an $n$-plet $\psi$ (where we limit ourselves to the $n=3,\,4$ or $5$ cases), which allows for electroweak-scale dark matter masses. Both $\chi$ and $\psi$ are odd under a discrete $\mathbb{Z}_2$ symmetry which is exclusive to the dark matter sector, stabilising the lightest mass eigenstate. They are coupled via higher-dimensional operators with a suppression scale $\gtrsim$ TeV. The relic density is very efficiently depleted via coannihilation, provided that the $\chi$ and $\psi$ masses are similar and that $\chi$ is in equilibrium with the thermal bath until freeze-out. It was shown in \cite{Bharucha:2017ltz} that, for suitable masses and Wilson coefficients, direct detection constraints can be avoided, while indirect detection is not sensitive to this type of model.

The models of \cite{Bharucha:2017ltz} are not UV complete, but constitute effective field theories suitable for describing the physics relevant for the LHC. The aim of this paper is to demonstrate how present and future LHC searches can be used to probe them. To this end we mostly focus on the case $n=5$ since it provides some very interesting and rather exotic signatures,\footnote{For recent discussions of aspects of the collider phenomenology of fermionic quintuplets, see e.g.~\cite{Ostdiek:2015aga, DiLuzio:2015oha, Lopez-Honorez:2017ora, Agarwalla:2018lnq}.} but we will also study the $n=3$ model and comment on $n=4$.  The case of $n=3$ is well-studied either as ``bino-wino'' dark matter, for which extensive literature on possible LHC searches already exists (see e.g.~\cite{Nagata:2015pra, ArkaniHamed:2006mb, Ibe:2013pua, Baer:2005jq, Rolbiecki:2015gsa}), or in other forms (see \cite{Chardonnet:1993wd} for an early model).

A fermionic $\SU{2}$ $5$-plet $\psi$ with zero hypercharge contains a doubly charged and a singly charged Dirac fermion $\psi^{\pm\pm}$ and $\psi^\pm$. Pairs of doubly-charged fermions will be copiously produced in the Drell-Yan process. They will decay in two-step cascades into dark matter and Standard Model states, with a decay length that can be macroscopic. We therefore propose the use of searches for displaced leptons (i.e.~leptons with a non-zero impact parameter with respect to the interaction vertex) as a means of probing this model, and possibly also related models involving higher $\SU{2}$ representations and similar properties. The $n=3$ model being similar to the bino-wino sector of the Minimal Supersymmetric Standard Model is constrained by soft lepton searches originally used to constrain supersymmetric neutralinos and charginos. Some corners of the parameter space of our models give rise to charged particles which are stable or quasi-stable on collider scales, and are hence largely ruled out already. 

In summary, our models constitute well-motivated frameworks in which long-lived particles feature naturally. They can be regarded as a testing ground for many of the ideas which have surfaced recently on the subject of long-lived particle searches at the LHC (see e.g.~\cite{Brooijmans:2018xbu} and references therein).

\section{Models}

The particle content of what we call next-to-minimal dark matter, or well-tempered $n$-plet dark matter, is that of the Standard Model plus a Majorana singlet $\chi$ and a fermionic $\SU{2}$ $n$-plet $\psi$, where $n\geq 3$. For $n$ odd and in particular $n=3$ or $n=5$, which are the cases we will be concerned with, $\psi$ is a Majorana fermion with hypercharge $0$, and the Lagrangian is
\begin{eqnarray}
 {\cal L} & =  & {\cal L}_{\rm SM}+ i\,\psi^\dag\ol\sigma^\mu D_\mu\psi+i\,\chi^\dag\ol\sigma^\mu\partial_\mu\chi \nonumber \\
 & & -\left(\frac{1}{2} M\psi\psi+\frac{1}{2}m\chi\chi\hc\right)+\text{(higher-dimensional operators)}
\end{eqnarray}
We assume both mass parameters $m$ and $M$ to be real and of the order of the electroweak scale, between about 100 and 500 GeV, and that $m < M$. A $\mathbb{Z}_2$ symmetry ensures the stability of the lightest dark matter sector state which on the renormalizable level is purely given by $\chi$, completely sterile, and therefore not a good thermal dark matter candidate. However, $\chi$ can interact with the Standard Model states and with $\psi$ through higher-dimensional operators, allowing for coannihilation with $\psi$ to deplete the $\chi$ abundance and to thus reproduce the observed dark matter relic density \cite{Bharucha:2017ltz}, $\Omega h^2=0.1188\pm 0.0010$~\cite{Ade:2015xua}. 

\subsection{The quintuplet model}
In the quintuplet case where $n=5$, $\psi$ transforms in the $\mathbf{5}_0$ of $\SU{2}_L\times\U{1}_Y$. The leading operator coupling $\chi$ to $\psi$ and the Standard Model Higgs doublet $\phi$ arises at dimension 7,
\be\label{eq:mixing}
{\cal O}_{\rm mix} = \frac{1}{\Lambda^3}\;(\phi^{\dag}\phi\phi^{\dag}\phi) \psi \chi\hc
\ee
The parentheses in this and the following equations are used to indicate the contraction of $\SU{2}$ indices to form a single irreducible representation, in that case the contraction of four Higgs doublets into a hypercharge-neutral state of weak isospin 2. The leading coupling of the $\psi\chi$ bilinear to the $Z$ boson arises at dimension 8, but despite its effect being suppressed by $\left(\frac{\vev{\phi}}{\Lambda}\right)^4$, it may nevertheless play a role in the decays of the $\psi$-like mass eigenstates:
\be\label{eq:Zcoupling}
{\cal O}_{\psi\chi Z}=\frac{1}{\Lambda^4}\;(\phi^\dag\psi^\dag)\ol\sigma^\mu D_\mu(\chi\phi\phi^\dag\phi)\hc
\ee

We assume that the suppression scale of such higher-dimensional operators is $\Lambda\gtrsim 1$ TeV, in order for an effective field theory description to be under control. We also assume that any four-fermion operators between $\chi$ and SM fermions $f$, $f'$ such as $f\chi f'{}^\dag\chi^\dag$ are suppressed, since these would otherwise induce (generically large) flavour-changing neutral currents. This can be phrased in terms of allowed $B$ and $L$ charges for the states that were integrated out at the scale $\Lambda$. Moreover, the Wilson coefficient of the dimension-5 operator $\chi\chi\phi^\dag\phi$ is constrained by direct detection experiments.\footnote{See \cite{Bharucha:2017ltz} for a more detailed discussion of higher-dimensional operators in this model and their impact on the relic density and on direct detection.}
The model is not UV complete, but contains all the states relevant for dark matter and (present-day) collider observables.

The mass eigenstates are a doubly charged Dirac fermion $\psi^{++}$, a singly charged Dirac fermion $\psi^+$ and two neutral Majorana fermions $\chi_{1,2}$, the heavier of which is $\psi$-like. Since the details of the mass spectrum have a considerable impact on the phenomenology, we will delineate them in the following.

Concerning the $\psi$-like mass eigenstates, their masses are exactly $M$ when neglecting both higher-dimensional operators and loop effects. Electroweak corrections split up the degeneracy, giving, at the one-loop level,
\be\label{Deltap1l}
\Delta m_{\chp-\chi_2}^{\text{one loop}} = \frac{g^2}{16\pi^2}M\left(f\left(\frac{m_W}{M}\right)-c_w^2\,f\left(\frac{m_Z}{M}\right)\right)
\ee
and
\be\label{Deltapp1l}
\Delta m_{\psi^{++}-\chp}^{\text{one loop}}=3\;\Delta m_{{\psi^+}-{\chi_2}}^{\text{one loop}}\,,
\ee
where
\be
f(x)=\frac{x}{2}\left(2\,x^3\log x-2\,x+\sqrt{x^2-4}(x^2+2)\log\frac{x^2-x\sqrt{x^2-4}-2}{2}\right)\,.
\ee
Numerically the loop-induced mass splittings are around $160$ and $480$ MeV respectively, with a weak (logarithmic) dependence on $M$. 

These radiative mass splittings compete with tree-level mass differences induced by higher-dimensional operators. The operator
\be
{\cal O}_{\psi\psi,5}=\frac{1}{\Lambda}(\psi\psi)(\phi^\dag\phi)\hc
\ee
will shift all the masses of the $\psi$-like states by the same amount. This effect can therefore be absorbed in a redefinition of $M$. A na\"ive dimension-5 candidate operator for isospin-dependent mass shifts,
\be
{\cal O}'_{\psi\psi,5}=\frac{1}{\Lambda}(\psi T^a\psi)(\phi^\dag\tau^a\phi)\hc
\ee
(where the $T^a$ and $\tau^a$ are the generators of the quintuplet and fundamental representation respectively) vanishes identically since ${\bf 5}\otimes{\bf 5}\supset{\bf 3}_a$, and the antisymmetric contraction of two $\psi$s is zero. The leading operator for non-universal mass shifts thus appears at dimension 7, involving a completely symmetric $\SU{2}$ tensor $C^{ABC}$ \cite{Bharucha:2017ltz} is
\be\label{eq:Deltahdo}
{\cal O}_{\psi\psi,7}=\frac{1}{\Lambda^3}C^{ABC} \psi^A\psi^B(\phi^\dag\phi\phi^\dag\phi)^C\hc
\ee
where $(\phi^\dag\phi\phi^\dag\phi)^C$ is the contraction of the four Higgs doublets to form a quintuplet. For $\Lambda\gtrsim 1$ TeV and an ${\cal O}(1)$ Wilson coefficient, this operator will give a $\lesssim (1$ GeV) contribution to the mass differences between $\psi^{++}$ and $\psi^+$ as well as between $\psi^+$ and $\chi_2$, of the same order as the calculable loop-induced mass differences of Eqs.~\eqref{Deltap1l} and \eqref{Deltapp1l}. However, for a smaller Wilson coefficient the contribution of Eq.~\eqref{eq:Deltahdo} may well become subdominant. 

The impact on the mass spectrum of the effective quintuplet-singlet mass mixing operator, obtained by replacing all Higgs fields by their vacuum expectation values $\vev{\phi}=v=174$ GeV in Eq.~\eqref{eq:mixing}, is negligible: Parametrically it is given by
\be
\Delta m_{\chp-\chi_2}^{\text{mixing}}\sim\theta^2(M-m)
\ee
where 
\be\label{eq:theta}
\theta=\sqrt{\frac{2}{3}}\frac{v^4}{\Lambda^3(M-m)}
\ee 
is the mixing angle.

Requiring that $\chi_1$ be a good dark matter candidate, with the observed relic density obtained through coannihilation with $\psi$, implies that $\frac{M-m}{M}$ should be of the order of $10\%$ or more precisely between 15--50 GeV \cite{Bharucha:2017ltz}. Any subdominant corrections to this quantity can always be absorbed in a redefinition of $m$.

Note that, since the operator of Eq.~\eqref{eq:mixing} giving rise to quintuplet-singlet mixing is dimension-7, demanding $\Lambda\gtrsim 1$ TeV results in a mixing angle $\theta$ which is generically very small according to Eq.~\eqref{eq:theta}. More precisely, for $M-m=15$ GeV, $\Lambda=1$ TeV and a Wilson coefficient normalised to $1$ one obtains $\theta \approx 0.05$ as an approximate upper bound, with much smaller values clearly possible. We estimate a lower bound on $\theta$ based on the dark matter production mechanism in Section \ref{sec:mixing}.

To summarise, in the well-tempered quintuplet model the new particles are the three $\psi$-like states $\psi^{++}$, $\psi^+$ and $\chi_2$ which are mass-degenerate at the sub-percent level, and the $\chi$-like state $\chi_1$ which is lighter by a few tens of GeV. The mass degeneracy between the charged mass eigenstates will play a crucial role for phenomenology. 

\subsection{The triplet model}

In the $n=3$ case the coannihilation partner $\psi$ transforms in the $\mathbf{3}_0$ of $\SU{2}_L\times\U{1}_Y$. The particle spectrum of the triplet model comprises a singly charged mass eigenstate $\chp$ and two neutral states $\chi_2$ and $\chi_1$. The $\psi$-like neutral and charged states are again split by a dimension-7 operator similar to that of Eq.~\eqref{eq:Deltahdo}, as well as by electroweak loops. They are therefore approximately mass-degenerate, with the radiative splitting given by Eq.~\eqref{Deltap1l}. We remark that, in the triplet model, mixing between the triplet and the singlet is induced already at dimension $5$ and an effective $\chi\psi Z_\mu$ vertex arises at dimension 6:
\be\label{eq:mixingtrip}
{\cal O}_{\rm mix}=\frac{1}{\Lambda}\;(\phi^{\dag}\phi) \psi \chi\hc
\ee
\be\label{eq:Zcouplingtrip}
{\cal O}_{\psi\chi Z}=\frac{1}{\Lambda^2}\;(\phi^\dag\psi^\dag)\ol\sigma^\mu D_\mu(\chi\phi)\hc
\ee
Thus, the cut-off scale can be moved to very large values while $\chi_1$ remains sufficiently strongly coupled to be a thermal dark matter candidate (see section \ref{sec:mixing}). Explicitly, the mixing angle in the triplet model following from Eq.~\eqref{eq:mixingtrip} ,with a Wilson coefficient normalized to unity, is given by
\be
\theta=\sqrt{2}\frac{v^2}{\Lambda(M-m)}\,.
\ee

\subsection{The quadruplet model}

The gauge eigenstates in the quadruplet model consist of a Dirac fermion $(\psi,\ol\psi)$ in the ${\bf 4}_{1/2}\oplus\ol{\bf 4}_{-1/2}$ (a variant would be the hypercharge $3/2$ case which also contains an electrically neutral state) besides the Majorana singlet $\chi$. They mix through the two independent dimension-6 operators
\be
\frac{1}{\Lambda^2}\ol\psi\chi\phi^\dag\phi\phi^\dag\,,\qquad \frac{1}{\Lambda^2}\psi\chi\phi\phi^\dag\phi\,.
\ee 
There are three neutral, two singly charged and one doubly charged mass eigenstates. Unlike the $n=5$ model, the tree-level mass splittings between the $\psi$-like states are already induced at dimension 5 through the operators
\be\label{eq:4pletsplittings}
\frac{1}{\Lambda}(\phi^\dag\tau^a\phi)(\ol\psi t^a\psi)\,,\quad \frac{1}{\Lambda}(\phi_i \tau^a{}^i_j\phi^j)(\ol\psi_I t^a{}_J^I\ol\psi^J)\,,\quad
\frac{1}{\Lambda}(\phi^\dag_i\tau^a{}^i_j\phi^{\dag j})(\psi_I t^a{}_J^I\psi^J)\,.\\
\ee
Here doublet indices $i,j$ are raised and lowered with the epsilon symbol, and quadruplet indices $I,J$ with the tensor $\sigma^1\tensor i\sigma^2$.
The isospin-dependent mass splittings of Eq.~\eqref{eq:4pletsplittings} are of the order of $\frac{v^2}{\Lambda}\sim {\cal O}(10)$ GeV for a ${\cal O}(1)$ Wilson coefficient and a cut-off scale in the TeV range. Contrarily to the quintuplet case, these mass splittings  are therefore not particularly small, unless the corresponding Wilson coefficients are suppressed or the cut-off scale is much higher, and generically dominate over the radiative contributions. Because of the abundance of states and free parameters, the model is much less predictive than the quintuplet model, and signatures from long-lived states are expected only in a small part of the overall parameter space.

\section{Estimate for the minimal mixing angle}
\label{sec:mixing}

In our models of well-tempered $n$-plet dark matter, the $\psi$-like states act as coannihilation partners for the dark matter particle $\chi_1$. The mixing angle between the singlet-like and the $n$-plet-like neutral mass eigenstates can be rather small, in which case both $\chi_1-\chi_1$ annihilation and $\chi_1-\psi$ annihilation are inefficient. Instead, the relic density of $\chi_1$ is depleted by the scattering and inverse decays $\chi_1 X\into \psi X'$, $\chi_1\into \psi X(X')$, where $X, X'$ denote suitable SM states and $\psi$ denotes any of the $\psi$-like states $\chi_2$, $\psi^\pm$ or $\psi^{\pm\pm}$. The standard formalism for coannihilation will give reliable results for the dark matter relic density, provided that the mixing angle is still large enough for $\chi_1$ to remain in chemical equilibrium until $\psi$ annihilation freezes out.

Recent studies \cite{DAgnolo:2017dbv, Garny:2017rxs} have pointed out that even smaller couplings between dark matter and its coannihilation partners may still give rise to a calculable thermal relic abundance. However, if the processes depleting the $\chi$ number density become inefficient (or close to inefficient) before $\psi$-$\psi$ annihilation freezes out, the resulting relic density significantly deviates from that predicted by standard coannihilation. The observed dark matter abundance will only be reproduced when changing the other model parameters, such as the mass difference between $\chi$ and $\psi$, with respect to the coannihilation scenario. It is therefore of interest to estimate the minimal mixing angle for which the standard coannihilation prediction is still valid, and our assumptions for the mass difference $M-m$ generating the observed relic density hold.  We emphasize that for smaller values of the mixing angle the model is still viable, but the formalism of \cite{DAgnolo:2017dbv, Garny:2017rxs} needs to be used to calculate the relic density.\footnote{For some even smaller value the dark matter candidate would effectively decouple from the thermal bath already at very large temperatures and overclose the universe, but this part of the parameter space is ruled out by disappearing track searches, as will be detailed below.}

In the triplet and quintuplet models, in the limit of a large cut-off scale $\Lambda$ and a correspondingly small mixing angle, the two effective vertices responsible for $\chi_1$ conversion into $\psi$ states are $h\chi_1\chi_2$ and $\chi_1\chpm W^\mp_\mu$ (the $\chi_1\chi_2 Z_\mu$ effective vertex being suppressed by a further factor of $v/\Lambda$). The SM states $X$, $X'$ involved in $\chi_1 X\,\leftrightarrow\,\psi X'$ scattering are therefore $W^\pm,\,\gamma,\,h$ and SM fermions. We assume that, at some early time, $\chi_1$ was in chemical equilibrium. With the further simplifying assumptions that
\begin{itemize}
 \item kinetic equilibrium is maintained for all states (and in particular for $\chi_1$ due to inelastic scattering processes),
 \item both the $\chpm$ and the $\chi_2$ number densities are given by their equilibrium densities until $x_f\approx 25$, since the $n$-plet-like state annihilate efficiently,
 \item 2-body decays $\psi^\pm\to\chi_1 W^\pm$ and $\chi_2\to\chi_1 h$ are kinematically forbidden,
 \item both $\chi\chi$ annihilation and $\chi\psi$ annihilation are negligible,
 \item decays and inverse decays are subdominant with respect to $2\into 2$ processes
\end{itemize}
\noindent (all of which are well justified for the parameter range we are considering) we can write down the Boltzmann equation governing the number density for $\chi_1$:
\be
\frac{\d}{\d x} Y_{\chi_1}=-\frac{1}{Hx}\left(\frac{M}{m}\right)^{3/2}\,e^{-\frac{M-m}{m}x}\left(g_{\chi_2}\Gamma_{\chi_2\into\chi_1}+g_{\chpm}\Gamma_{\chpm\into\chi_1}
\right)\left(Y_{\chi_1}-Y_{\chi_1}^{\rm eq}\right)\,.
\ee
Here
\be
H=\sqrt{\frac{\pi^2 g_*(x)}{90}}\frac{m^2}{M_P}\frac{1}{x^2}
\ee
is the Hubble parameter, $g_{\chi_1}=g_{\chi_2}=2$, $g_\chpm=4$, $g_*$ is the effective number of degrees of freedom, $Y_{\chi_1}$ is the yield, and the thermally averaged conversion rates $\Gamma_{\chi_1\to\chi_2}$ and $\Gamma_{\psi^\pm\to\chi_1}$ are defined as in \cite{Garny:2017rxs}. We find that the dominant processes are those involving light SM fermions in the final and initial state.
On solving the Boltzmann equation, we estimate the minimum mixing angle by demanding that the $\chi_1$ abundance deviates from the equilibrium value by at most $10\%$ for $x=25$.
 
\begin{figure}
\begin{center}
\includegraphics[width=.45\textwidth]{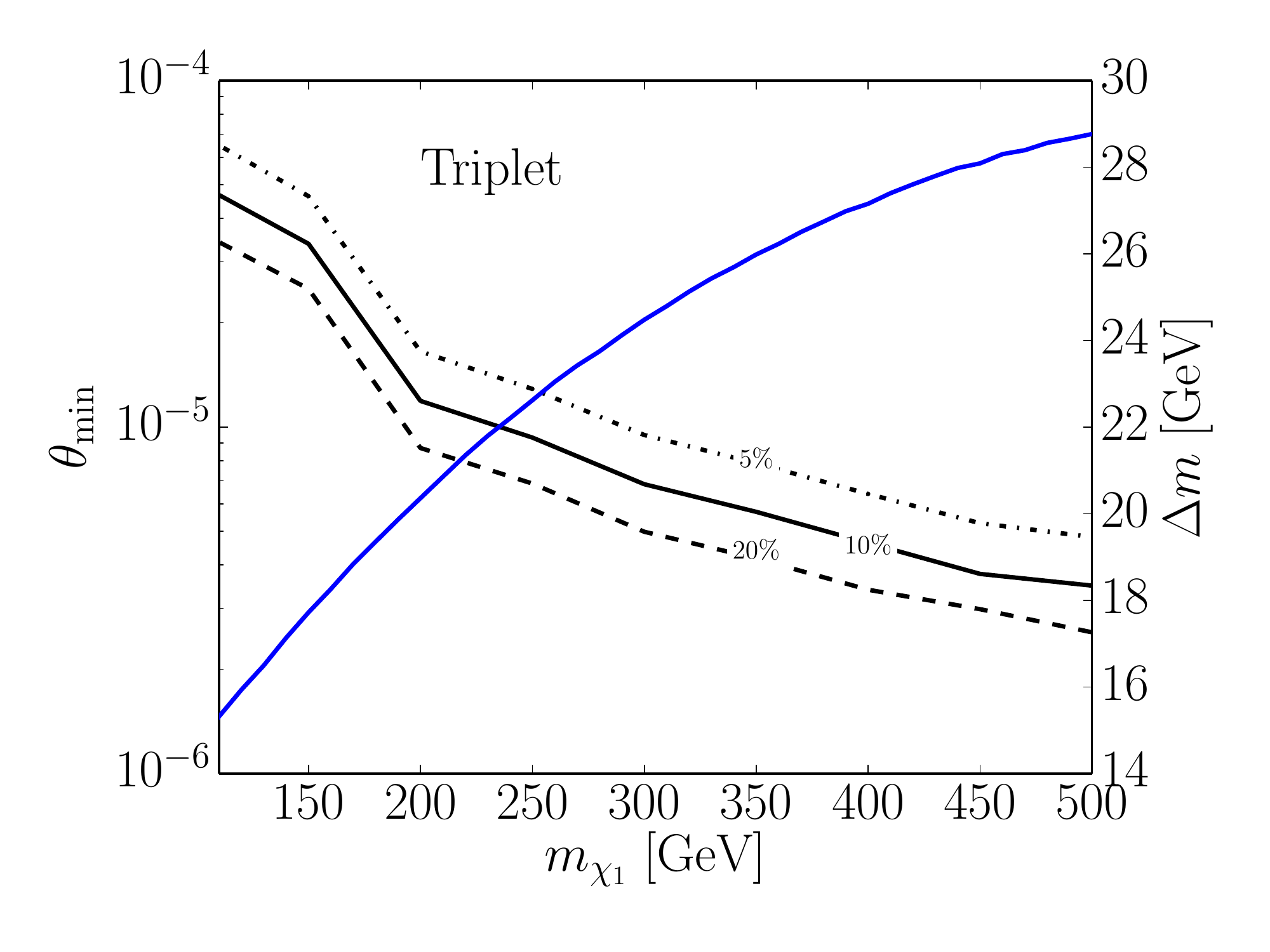}
\includegraphics[width=.45\textwidth]{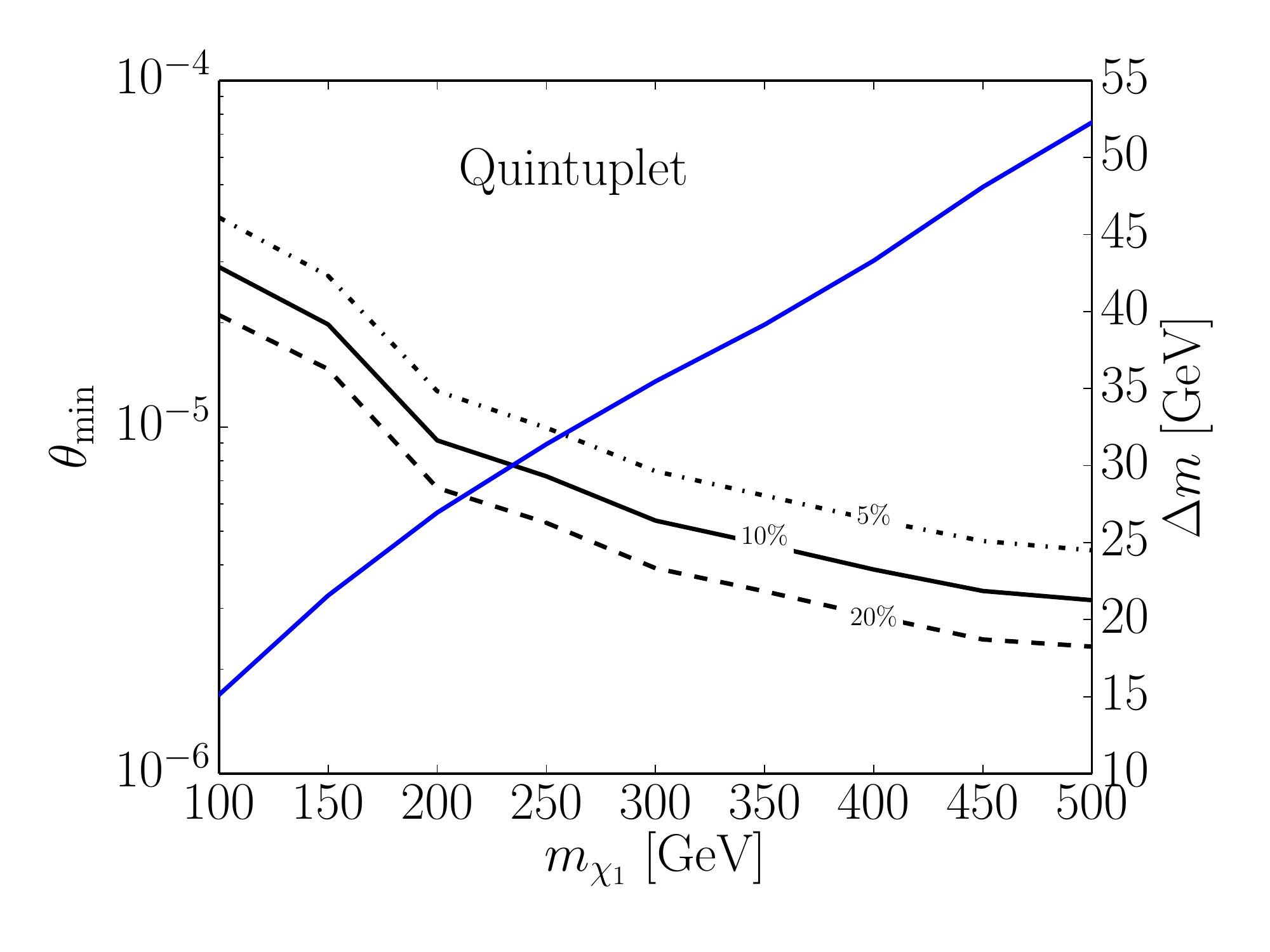}
\caption{\label{fig:min-theta} The minimal angle for the triplet (left) and quintuplet (right) case which it is appropriate to calculate the relic density via the freeze-out of the coannihilation process. The three black curves indicate the values of the mixing angle which result in a $5\%$ (dash-dotted), $10\%$ (solid) and $20\%$ (dashed) deviation of the $\chi_1$ number density from its equilibrium value as the coannihilation processes freeze out at $x_f\approx 25$. The blue curves show the predicted mass splitting between the n-plet and singlet masses to obtatin the correct relic density through coannihilation.}
\end{center}
\end{figure}

The resulting minimal angle as a function of $m_{\chi_1}$ is shown in Fig.~\ref{fig:min-theta}. In order to indicate the theory uncertainty, we also plot the values of the mixing angle when demanding at most a $5\%$ or a $20\%$ deviation of the $\chi_1$ number density from its equilibrium value at $x=25$.

\section{Collider Phenomenology}

Let us start by summarising the parameters relevant for collider phenomenology.  In the quintuplet model, aside from the masses of the particles, two mass differences are particularly important. These are $\Delta m = m_\chp - m_{\chi_1}$, which is fixed by requiring the observed relic density through coannihilation, and $\delta m = m_\chpp - m_\chp$, which determines the lifetime of the doubly charged state. A third important parameter is the mixing angle $\theta$ which, according to the previous section, should satisfy $10^{-(5-6)}\lesssim \theta\lesssim 0.05$. Fixing $\Delta m$ and $\theta$, and normalizing the Wilson coefficient of the mixing operator of Eq.~\eqref{eq:theta} to $1$, automatically sets the suppression scale $\Lambda$.  

When  higher-dimensional operators such as the one of Eq.~\eqref{eq:Deltahdo} are negligible, the quintuplet model becomes very predictive as the mass difference $\delta m$ between the doubly charged and singly charged states is uniquely fixed by electroweak loops. Moreover, in that case the only 2-body decay mode available for $\psi^{++}$ is $\psi^{++}\into\pi^+\psi^+$ with an ${\cal O}(1)$ branching fraction. The decay width of charged states in the isospin-$j$ representation of SU(2) is
\begin{align}
 \Gamma (\psi_{j_3+1} \rightarrow \psi_{j_3}) &  =  \frac{T_+^2 G_F^2 V_{ud}^2 \delta m^3 f_\pi^2}{\pi} \sqrt{1-\frac{m_{\pi^+}^2}{\delta m^2}} \\
\text{where }  T_+^2 &  =  j(j+1) - j_3(j_3+1)  = 4\,. \nonumber
\end{align} 
This gives
\begin{equation}
\Gamma \approx \delta m^3\; 2.8 \times 10^{-12} \gev^{-2}
\end{equation}
or, for $\delta m=\Delta m^{\text{one-loop}}_{\chpp-\chp}=480$ MeV,
\be
\Gamma=3.1\times 10^{-13}\gev\qquad\Rightarrow\qquad c\tau=0.6\text{ mm}\,.
\ee
This decay length is macroscopic, in the sense that it can potentially be resolved by ATLAS and CMS.

In general, there will be an unknown contribution to the mass splitting between the doubly charged and the singly charged fermion which depends on the UV completion. This contribution, induced by the operator of Eq.~\eqref{eq:Deltahdo}, is at most of the same order as the one-loop mass splitting, and can be either positive or negative. 
To account for this effect we will also study values for $\delta m$ which are different from the radiative one. As long as $140\mev\lesssim\delta m\lesssim 800\mev$ the phenomenology does not change much qualitatively, except for the varying decay length, since the $\chpp$ continues to decay predominantly into a single pion. For larger mass splittings the $\chpp$  decay is effectively prompt, while for $\delta m\lesssim m_{\pi^+}$, $\chpp$ becomes stable on detector scales.  

The decays of the other $\psi$-like states can proceed through the operators of Eq.~\eqref{eq:mixing} and \eqref{eq:Zcoupling}. The singly charged $\chp$ generically decays mostly into the lighter state $\chi_1$, either through its small mixing with the $\psi^0$ or via the dimension-8 coupling of Eq.~\eqref{eq:Zcoupling}. However, in the regime where the corresponding Wilson coefficients are too small (for mixing angles $\lesssim 10^{-5}$), the decay into $\chi_2$ will become important despite being phase-space suppressed. In that case the $\chp$ will likewise become long-lived on the scale of the detector.  A corresponding region of the parameter space also exists in the triplet model, where the relevant couplings arise from the operators of Eq.~\eqref{eq:mixingtrip} and \eqref{eq:Zcouplingtrip}. For a sufficiently large cut-off scale $\Lambda$, the overall decay width will become of the order of the partial width into $\chi_2$, corresponding to a decay length of $\approx 10$ cm characteristic for an approximately ``wino-like chargino'', leaving a seemingly disappearing charged track in the detector. 

The decay of the heavier $\psi$-like neutral state $\chi_2$ can proceed through either the dimension-7 coupling of Eq.~\eqref{eq:mixing} and an off-shell Higgs boson, or through the dimension-8 coupling of Eq.~\eqref{eq:Zcoupling} and an off-shell Z. The former decay modes are also suppressed by the masses of the Standard Model fermions in the final state (predominantly $m_b$), so either mode may be the dominant one. For sufficiently small mixing angles, the $\chi_2$ decay length can become macroscopic. The same applies to the triplet model.

The production modes that can possibly lead to sizeable signal at the LHC are:
\begin{itemize}
\item $\chpp \chmm$ \\
The LHC pair-production cross-section for $\psi^{++}$ is significant thanks to its large electric charge and weak isopsin. After production of a $\chpp \chmm$ pair, cascade decays $\psi^{\pm\pm}\into\psi^\pm\pi^\pm$, $\psi^\pm\into W^{*\pm}(\into \ell\bar\nu_\ell)\chi_1$ can lead to lepton pairs whose tracks are displaced (i.e. have a non zero transverse impact parameter) with respect to the primary vertex thanks to the long $\chpp$ lifetime. The other final state particles cannot be detected (the pions are so soft that they will be buried in low-energy QCD backgrounds, while the dark matter particles and neutrinos escape the detector unseen). However, displaced lepton pairs can be and are being searched for.

\item $\chpp \chm$ and $\chmm \chp$ \\
The primary search channel here would be dileptons, one of which would be displaced. This final state currently falls through selection criteria of both the prompt dilepton and displaced dilepton searches. The exchange of $W^\pm$ means there would be a measurable excess of the positively charged over negatively charged displaced leptons which cannot be obtained from the irreducible heavy flavour backgrounds.  The primary irreducible background is from heavy quarks and would have to be data-driven.

\item $\chp \chm$ \\
The production of singly-charged pair of fermions which decay to leptons or jets and missing energy is a long-established supersymmetric search and directly applicable to our model.  Depending on the effective couplings between the singlet and the $n$-plet state, the $\chp$ may be long-lived, thus giving either a displaced lepton or disappearing track signature. 

\item $\chp \chi_2$ + h.c.\\
The $\chi_2$ may decay via an off-shell Higgs boson to $b\bar b \chi_1$, or via an off-shell $Z$ boson. In the former case the signature would be a soft lepton with two soft b-jets and missing energy.  Due to very large background from $t\bar t$ production, this search does not have enough sensitivity to detect the production cross sections predicted by our model. In the latter case the most promising signature is again dileptons. 
There is a range of mixing where $\chp$ still decays within the detector but $\chi_2$ is stable on the scale of the detector, however a single lepton, even when accompanied by hard jets would be washed out by large SM background. 

\end{itemize}

\subsection{Displaced lepton searches}

For the quintuplet model, the largest production channel is $\chpp \chmm$. The produced states are long-lived, eventually decaying to the singly charged states which in turn decay to leptons displaced from the primary vertex. Not surprisingly, this turns out to be the most sensitive search channel for the 5-plet case, with the strongest limits from CMS search for a displaced electron-muon pair at 8 TeV with 20 fb$^{-1}$ \cite{Khachatryan:2014mea}. This search asks for an isolated electron and an isolated muon, both with a $p_T>25$ GeV, and a transverse impact parameter of at least 0.2 mm (``Signal Region (SR) 1''), 0.5 mm (SR2) or 1 mm (SR3). Due to the small mass splitting between the $\psi^\pm$ and $\chi_1$ states, the emitted leptons are soft such that the signal observed is limited by $p_T$ cuts on the leptons, which are optimised for a different model. Consequently a similar analysis at 13 TeV with 3 fb$^{-1}$ \cite{CMS:2016isf} is unfortunately not competitive because of the lepton $p_T$ cuts are tighter, 40 and 42 GeV for the muon and electron respectively, which is larger than the typical mass splitting expected from coannihilation and therefore leaving no signal events surviving.

\begin{figure}
\begin{center}
\includegraphics[scale=.5] {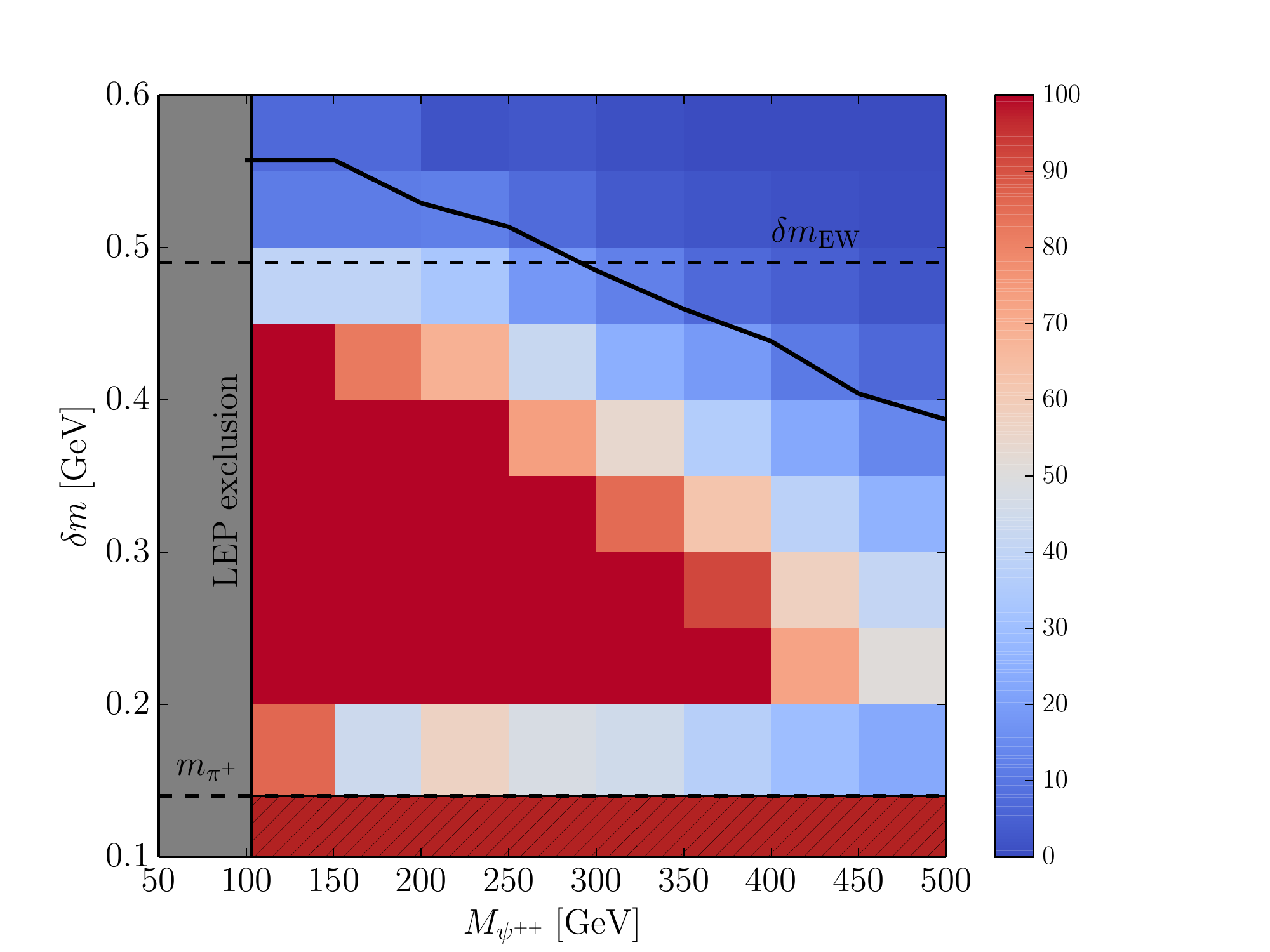}
\caption{\label{fig:excl-dm} Exclusions from the CMS displaced electron-muon pair search~\cite{Khachatryan:2014mea} in the $M$-$\delta m$ plane.  The dashed line corresponds to electroweak mass splitting of 0.49 GeV. The coloured heat-map shows the value of the test statistic $\mathcal{Q}$, defined in appendix~\ref{sec:validation}. The bottom hatched red area refers to limits from heavy stable charged track search whereas the grey vertical area refers to the LEP excluded region up to $m_{\psi^{++}} = 103$~GeV.}
\end{center}
\end{figure}

Details for the CMS search at 8 TeV with 20 fb$^{-1}$ and the validation of the reinterpretation are given in Appendix~\ref{sec:validation}. On reinterpreting the experimental exclusion bounds in terms of our model, we are able to exclude a region of parameter space in the $m$-$\delta m$ plane, which we display in Fig.~\ref{fig:excl-dm}.  In addition to the exclusion bound (solid black line), we indicate the radiative mass splitting of 0.49 GeV (dashed line), the value of the test statistic $\mathcal{Q}$, defined in Appendix~\ref{sec:validation} (coloured heat-map), limits from heavy stable charged track searches (red hatched region) and the LEP exclusion up to $m_\chpp = 103$~GeV (grey shaded region). For the case where the splitting between the charged states is only determined by electroweak corrections, we rule out $m_\chpp < 280$~GeV.  The falling shape of the curve is a consequence of the reduced cross section with increasing mass and increasing $d_0$ predicted by smaller values of $\delta m$.  The turning point is reached near $m_\chpp \sim 650$~GeV, however we choose to only present results for masses $<500$ GeV to ensure that an effective field theory description remains valid for low cutoff scales.  When  $\delta m$ falls below the pion mass, the resulting tracks are stable and ruled out by charged track searches which we examine more closely in the next section.

\subsection{Charged track searches}

One of the most striking characteristics of the quintuplet model is that the doubly charged states can be long lived, hence it can also be constrained by searches for heavy stable charged particles. The life-time can be particularly long if the corrections to the masses of $\chpp$, $\chp$ and $\chi_2$ due to the operator of Eq.~\eqref{eq:Deltahdo} dominate over the radiative corrections of Eqs.~\eqref{Deltap1l} and \eqref{Deltapp1l}, and if the corresponding Wilson coefficient is negative such that the mass difference between $\chpp$ and $\chp$ drops below the charged pion mass. In that case the $\chpp$ width will be dominated by the three-body decays $\chpp\into\chp e^+\bar\nu_e$ and possibly $\chpp\into\chp \mu^+\bar\nu_\mu$. Alternatively, the mass difference can be negative, making the $\chpp$ the lightest quintuplet-like particle and leaving open only five-body decay modes. 

In either of these cases, the doubly charged fermion becomes stable on collider scales. The most sensitive search to this scenario is a CMS search from 2016~\cite{CMS:2016ybj} for a Dirac fermion of hypercharge $0$ and weak isospin $2$. On reinterpreting the exclusion limits we find that this case, i.e.~when the Wilson coefficient is negative such that the $\chpp$ is stable on collider scales, is excluded for the entire mass range we are considering. 

\begin{figure}
\begin{center}
\includegraphics[width=.45\textwidth]{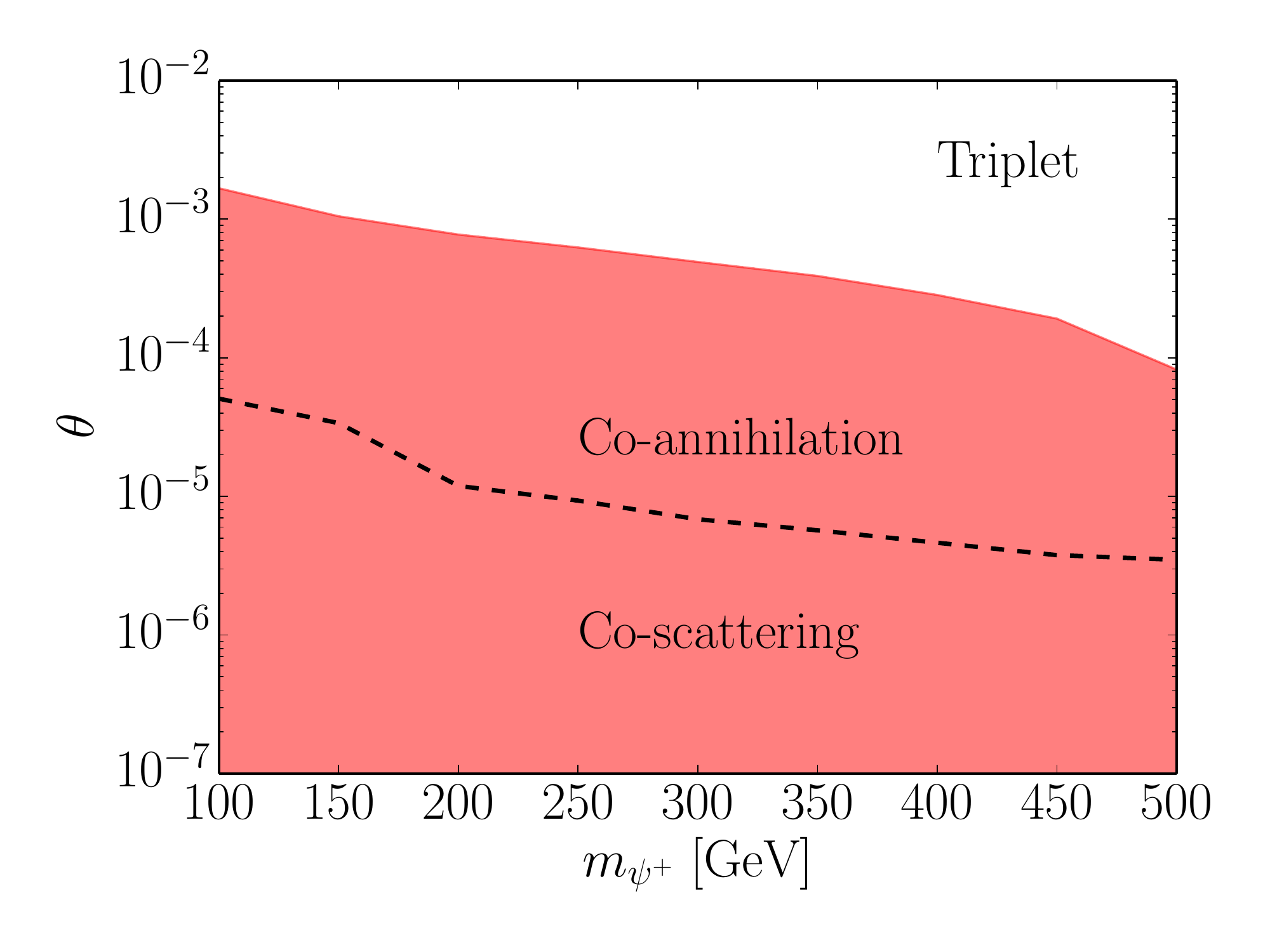}
\includegraphics[width=.45\textwidth]{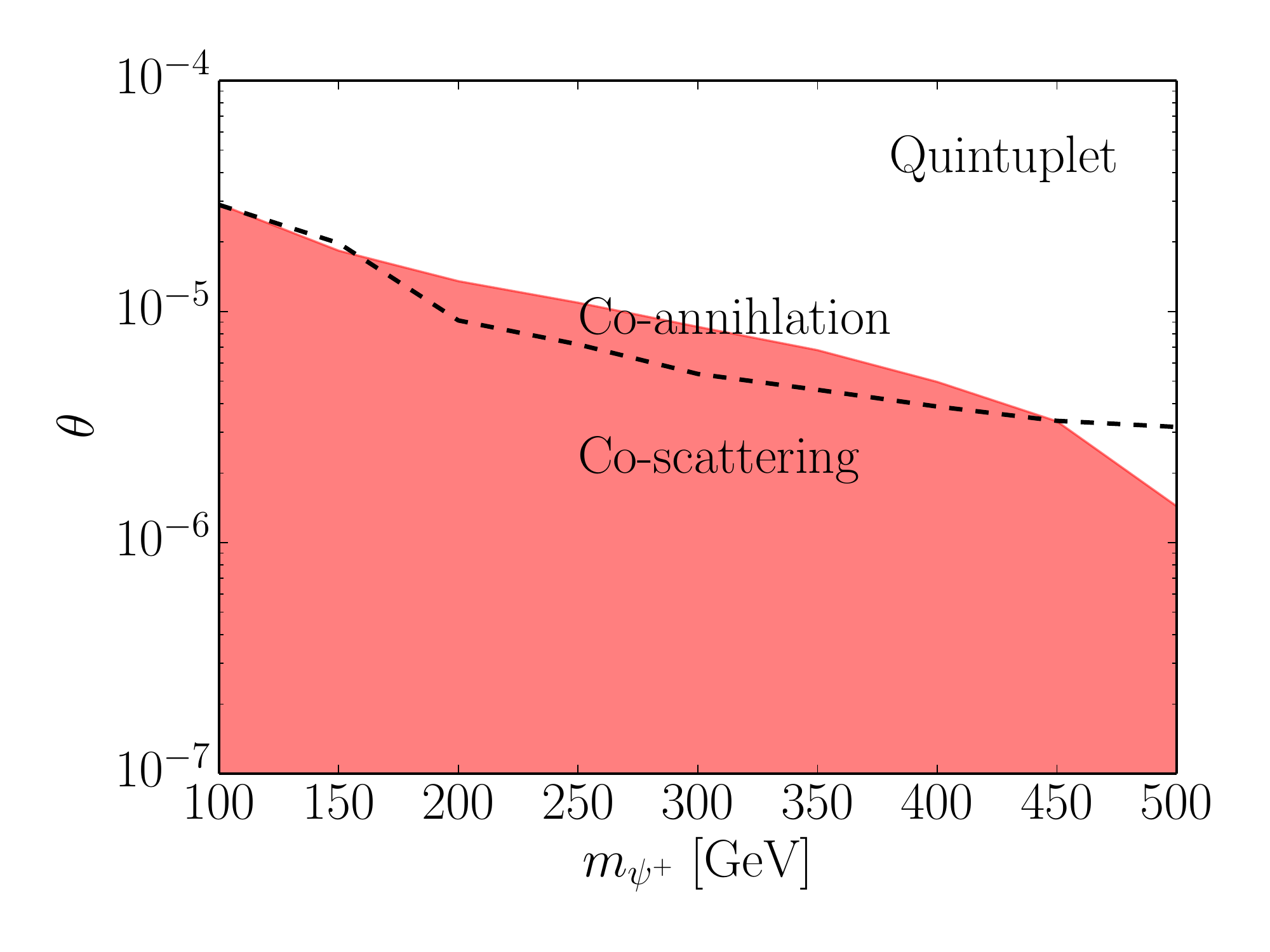}
\caption{\label{fig:diss-trk} The red shaded region shows exclusion from disappearing track search for the triplet (left) and quintuplet (right) interpreted in terms on the mixing angle $\theta$. The dotted line corresponds to minimum values of $\theta$ required by self-consistent calculation of relic density via coannihilation.}
\end{center}
\end{figure}

Both the quintuplet and the triplet model can give rise to long-lived singly charged states $\chp$ if the cutoff scale at which effective $\chp\chi_1 W$ couplings are mediated is sufficiently high or the corresponding Wilson coefficient is sufficiently small.  This happens for mixing angles $\theta \lesssim 10^{-5}$ for the quintuplet and $\theta \lesssim 10^{-3}$ for the triplet.  At $\theta\lesssim 10^{-6}$, the lifetime of the $\chp$ would be determined entirely by the mass difference between $\chp$ and $\chi_2$, which is 0.16 GeV at one-loop from EW corrections, similar to Winos in supersymmetry. The most sensitive search for this signature to date is the ATLAS analysis of  \cite{Aaboud:2017mpt}, which provides an interpretation in the context of Wino-like charginos and neutralinos in minimal anomaly-mediated supersymmetry breaking, and whose exclusion bounds therefore immediately apply to our triplet model. In the quintuplet, a further contribution to the signal can be from $\chpp$ decay discussed previously --- either when $\delta m$ is small giving doubly charged disappearing tracks or when $\delta m > 0.6$ which leads to prompt decays into long lived $\chp$.  However, here we only present results for the conservative assumption of direct production modes of the $\chp$. 

The exclusions on the mixing angle $\theta$, as a function of $m_{\psi^+}$, from the ATLAS disappearing track search on triplet and quintuplet model are shown in Fig.~\ref{fig:diss-trk}. For the triplet, these limits lie above the minimum mixing angle $\theta$ required for coannihilation calculated in section~\ref{sec:mixing}. Therefore, in all the remaining parameter space the relic density can be reliably calculated from coannihilation. For the quintuplet, we have also indicated the mixing angle where there is a transition between coannihilation and coscattering as the main mechanism governing the thermal relic density generation (see section \ref{sec:mixing}). At low and high masses, there are small ranges of mixing angles where coscattering starts becoming important and yet disappearing track searches are presently not sensitive.

In \cite{Ostdiek:2015aga} the disappearing track signature was used to recast the older analyses \cite{Aad:2013yna}, \cite{CMS:2014gxa} to a pure quintuplet model, excluding quintuplet masses of about 250 GeV with 20 fb$^{-1}$ of 8 TeV data.  The current reach of the disappearing track search at 13 TeV is $\sim 580$~GeV with 36 fb$^{-1}$ data.

\subsection{Searches for soft di-leptons}

In the triplet model, the main production channels are $\chp\chm$ and $\chp \chi_2$, both of which result in soft leptons due to the small splitting between the $\psi$-like and the $\chi_1$ states.
The most constraining analyses for the triplet model, at least in those parts of the parameter space where $\chp$ is not long-lived, therefore concern final states with soft leptons. In fact, the triplet model corresponds precisely to a well-tempered wino-bino system in supersymmetry, with all other superpartners parametrically heavy (and heavy higgsinos notably responsible for inducing singlet-triplet mixing at dimension 5 and couplings to the $Z$ boson at dimension 6), and can therefore be constrained in a straightforward manner by comparing with searches for supersymmetric charginos and neutralinos. 

In \cite{CMS:2017fij}, CMS presents a search for soft leptons at 13 TeV and 35.5 fb$^{-1}$ interpreted in a simplified model with some features of a bino-wino system. This analysis requires two isolated opposite-sign leptons of arbitrary (muon or electron) flavour, the leading of which should have transverse momentum between $5$  and $30$ GeV, in addition to a jet with $p_T>30$ GeV and missing transverse energy in excess of 125 GeV. An interpretation is provided in terms of supersymmetric charginos and neutralinos, where $\chi^0_2\chi^\pm_1$ production is assumed with a purely wino-like cross section. The $\chi^0_2$ and $\chi^\pm_1$ are assumed to subsequently  decay into $\chi^0_1$ and a virtual $Z$ and $W$ boson respectively. 

\begin{figure}[t]
\begin{center}
\includegraphics[scale=.5] {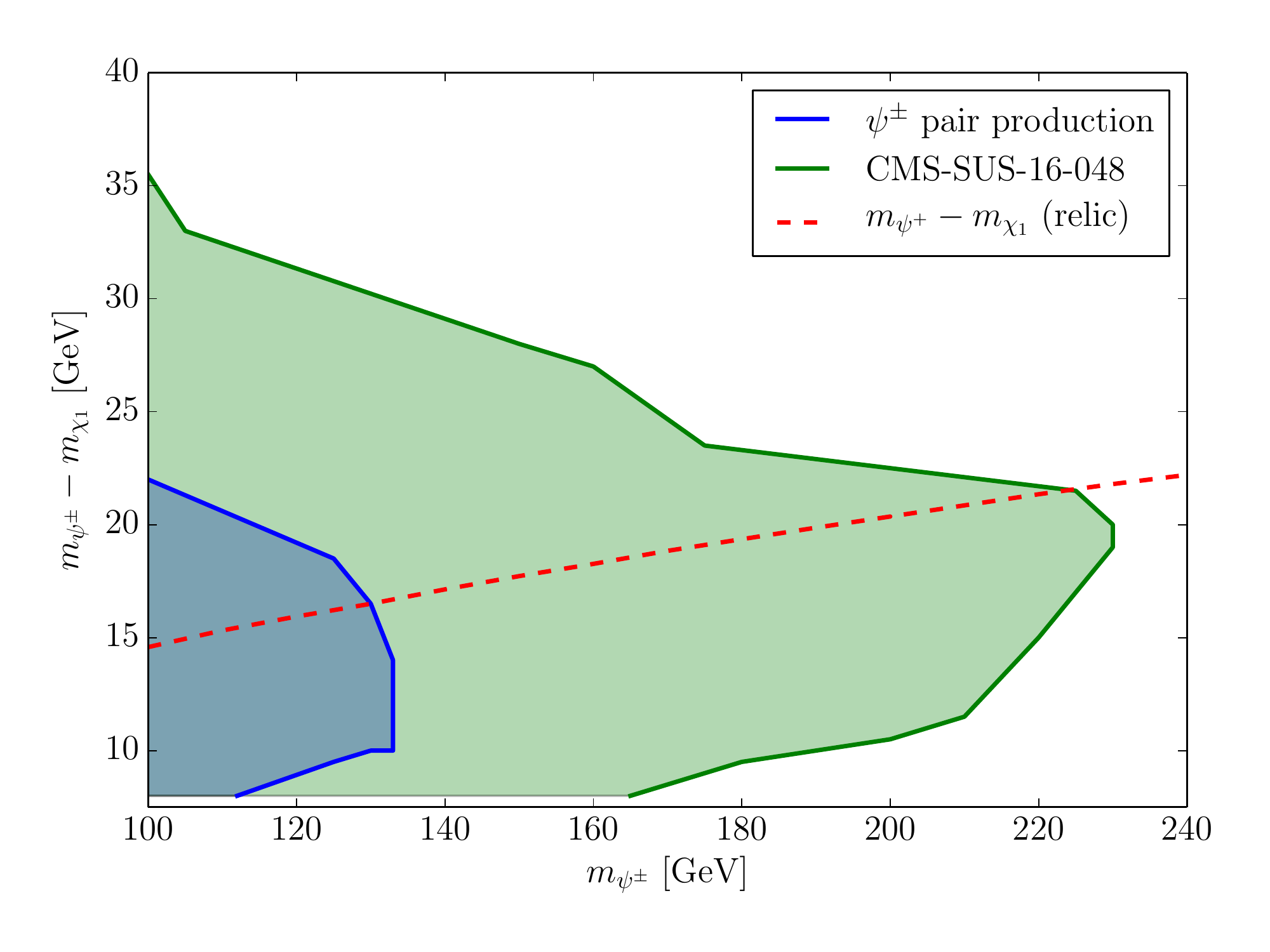}
\caption{\label{fig:excl-triplet} (Left panel) Exclusions from the CMS dilepton search \cite{CMS:2017fij} interpreted in the triplet model. Blue curve: only $\chp\chm$ pair production contributes to the dilepton signal, for $\chi_2$ decaying predominantly as $\chi_2\into\chi_1 h^*$. Green curve taken from \cite{CMS:2017fij}: only $\chi_2\chpm$ production contributes, assuming that $\chi_2$ decays predominantly as $\chi_2\into\chi_1 Z^*$. The dashed red line shows the predicted mass separation from requiring relic density of $\Omega h^2 = 0.12$.}
\end{center}
\end{figure}

These assumptions are not in exact correspondence with the true bino-wino (or equivalently triplet-singlet) model. Notably, $\chp\chm$ production is neglected, even though it can also contribute to a soft dilepton signal. Moreover, a significant fraction of the $\chi_2$ can decay through a virtual Higgs boson, in which case the leptonic final states have negligible branching fractions. Fig.~\ref{fig:excl-triplet} shows the results of the study \cite{CMS:2017fij} reinterpreted for a model in which the leptonic $\chi_2$ decays are negligible, such that any dilepton signal must originate from $\chp\chm$ production. For comparison, we also show the exclusion curve of \cite{CMS:2017fij}, corresponding to a negligible fraction of dileptons coming from $\chp\chm$ production and to $\text{BR}(\chi_2\into\chi_1 Z^*)=100\%$.  The shape of the curve is determined by two components --- the production cross section and efficiency of the $p_T$ cuts.  Looking at the green region for concreteness, the top of the curve is determined primarily by the falling production cross section will mass. We see that at the maximal value of 230 GeV, the reduced efficiency of the $p_T$ cuts dominates when the mass splitting falls below $\sim 20$~GeV.  In the case of pure $\psi^+ \psi^-$ production, this point is reached at 130 GeV.  Realistic models will be somewhere in between these two, depending on the relative importance of the operators which provide the leading-order coupling with the Higgs boson Eq.~\eqref{eq:mixingtrip} and the $Z$ boson Eq.~\eqref{eq:Zcouplingtrip}.

The quintuplet model can also be visible in the soft dilepton channel via $\chp\chm$ production, for which the limits are identical to the triplet case.  However limits from the displaced lepton search provide stronger constraints except in case of non zero contributions from operator in Eq.~\eqref{eq:Deltahdo} which render the $\psi^{++}$ decay prompt. 

\section{Conclusions}

Next-to-minimal dark matter is a viable and economical model of electroweak-scale dark matter with characteristic collider signatures. The dark matter sector consists of a fermionic singlet and its coannihilation partner, a fermionic $\SU{2}$ $n$-plet with $n\gtrsim 3$. These states interact via higher-dimensional operators, hence the model is not UV complete but serves as an effective field theory which is valid at energy scales relevant to the LHC. For $n=3$ and $n=5$ in particular, the model is very predictive, since the mass difference between the singlet and the $n$-plet is fixed by the relic density constraint. Its LHC physics then depends mostly only on two parameters, the overall mass scale and the singlet-$n$-plet mixing angle.

We have examined the bounds on these models from LHC searches, focussing on the $n=3$ (triplet) and $n=5$ (quintuplet) models.  For a dark matter mass in the range of $100-500$~GeV, the mass splitting with the coannihilation partner is of the order of $\sim 10-30$~GeV.  Therefore, these models often fail to satisfy hard $p_T$ cuts demanded by BSM searches.  For the triplet model, the best limits come from soft dilepton searches, i.e.~pair production of coannihilation partners and subsequent decays into leptons. Depending on whether the dominant decay channel of the neutral triplet-like state proceeds via an off-shell Higgs boson or an off-shell $Z$, we find that the limit on the triplet mass is between 130 and 230 GeV, the latter corresponding to the limit on supersymmetric wino-like charginos and neutralinos given by the experimental collaborations.

The quintuplet model is richer in its phenomenology as it contains doubly charged states which are naturally long-lived. Drell-Yan production of these states results in two oppositely charged leptons with large impact parameters which are detectable in the CMS displaced lepton searches.  We find that this is the channel with the highest sensitivity, and that the 8 TeV displaced lepton search currently places a limit of 280 GeV on the quintuplet mass. Unfortunately, the equivalent 13 TeV search is not sensitive to our model since the $p_T$ cuts have been tightened with respect to the 8 TeV analysis, which allows the background to be drastically reduced but in our case also leaves no signal events. For future analyses of this kind, we would urge the experimental collaborations to consider using kinematic cuts which remain sensitive to the $p_T$ range typical for mass splittings in coannihilation models, i.e.~around $10 - 30$ GeV.

We have also assessed the possibility of our models to produce disappearing track signatures. To this end we have estimated the minimal mixing angle $\theta$ between the $n$-plet and the singlet such that the latter remains in chemical equilibrium until $n$-plet annihilation freezes out. We obtain a minimal mixing angle of the order of $10^{-5}$ for both the quintuplet and triplet; for lower mixing angles the relic density would deviate from the value predicted by coannihilation. However, around this critical value of $\theta$, the lifetime of the singly charged $n$-plet state becomes macroscopic.  Disappearing track searches provide experiental lower bounds on $\theta$, ruling out mixing angles up to $\sim 1.1 \times  10^{-3}$ for the triplet and up to $\sim 3 \times 10^{-5}$ for the quintuplet model. These bounds are complementary to upper bounds from direct detection experiments which are currently only sensitive to values down to $\theta \sim 0.1$~\cite{Bharucha:2017ltz}.

\section*{Acknowledgements}

The authors thank Jamie Antonelli and Mathias Garny for correspondence, and Ronan Ruffault for contributing to this project at an early stage. This work has been carried out thanks to the support of the OCEVU Labex (ANR-11-LABX-0060) and the A*MIDEX project (ANR-11-IDEX-0001-02) funded by the "Investissements d'Avenir" French government program managed by the ANR.

\appendix

\section{Validation of CMS displaced lepton search.}
\label{sec:validation}

The CMS experiment has two displaced lepton searches ($\sqrt s = 8$ TeV, $\mathcal{L} = 19.7 \pm 0.5 \mathrm{fb}^{-1}$ \cite{Khachatryan:2014mea} and 
$\sqrt s = 13$ TeV, $\mathcal{L} = 2.6 \mathrm{fb}^{-1}$) \cite{CMS:2016isf}) that require oppositely charged, $e$ and $\mu$ with large impact parameters with respect to the primary vertex.  The second of these requires large $p_T$ from the leptons; given the highly compressed spectrum, it therefore has less reach in mass with the luminosity used compared to the 8 TeV search.  We therefore focus here on the 8 TeV search only.

The initial selection cuts are:
\begin{enumerate}
\item Select events with one $e$, one $\mu$, oppositely charged
\item Require $p_T > 25$ GeV and $\Delta R_{e\mu} > 0.5 $
\item For each jet (anti-kt, $R=0.5$, $p_T^{\text{min}} = 10$ GeV), require $\Delta R_{\ell j} > 0.5$
\item Transverse impact parameter $d_0 > 0.1 \text{mm}$.
\end{enumerate}

\begin{figure}[ht]
\begin{center}
\includegraphics[scale=.5] {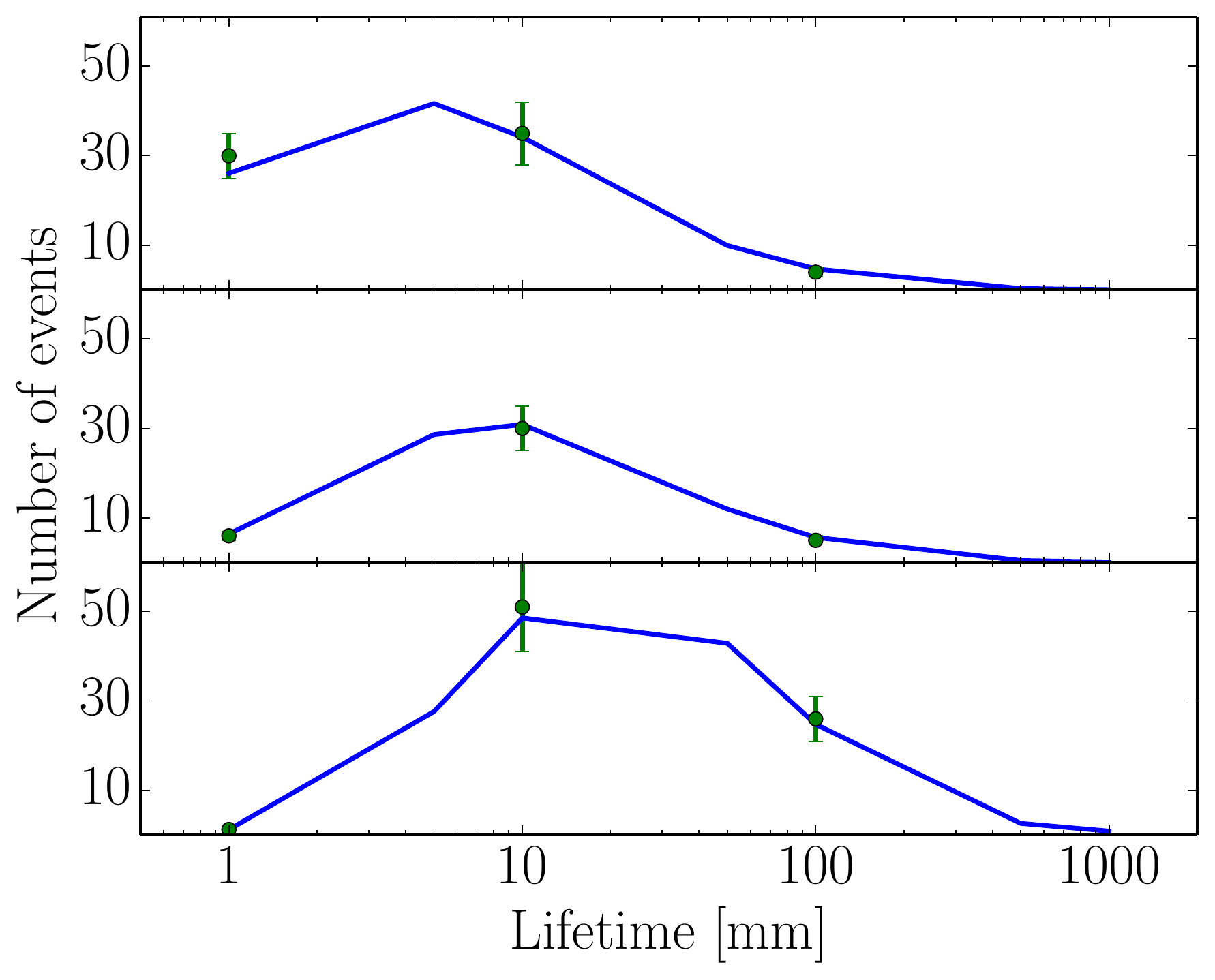}
\caption{\label{fig:excl-stop} Comparison of our expected signal (blue line) for the benchmark $M_{\tilde t} = 500$~GeV compared with published expected signal (green dots) in the three signal regions SR1 (top), SR2 (middle) and SR3 (bottom).}
\end{center}
\end{figure}

The comparison of our signal yields ( after applying efficiencies and cuts) with the published expectations for the CMS benchmark case is shown in Fig.~\ref{fig:excl-stop}.

For events passing the selection cuts, signal regions are defined as follows
\begin{description}
\item [\textsc{SR3:}] Both leptons satisfy  $1.0 \mm < d_0 < 20 \mm$.
\item [\textsc{SR2:}] One or both leptons fail \textsc{SR3} but satisfy $ d_0 > 0.5 \mm$
\item [\textsc{SR1:}] One or both leptons fail \textsc{SR2} but satisfy $ d_0 > 0.2 \mm$
\end{description}

The exclusion is determined based on a log-likelihood ratio of the signal-plus-background hypothesis with the background-only hypothesis.  The values for the expected background are taken from \cite{Khachatryan:2014mea}.  In each signal region $i$, given expected signal $s_i$, expected background $b_i$ and observed number of events $n_i$, we define the poisson likelihood as:

\begin{equation}
\mathcal{L}_{s+b} = \prod_{i=\text{SR1, SR2, SR3}} e^{-(s_i + b_i) } \frac{(s_i + b_i)^{n_i}}{n_i !}
\end{equation}

The test statistic $\mathcal{Q}$ is defined as 
\begin{eqnarray}
\mathcal{Q} & = &  - 2 \log \left( \frac{\mathcal{L}_{s+b}}{\mathcal{L}_{b}} \right) \nonumber \\
 & = & - 2 \sum_i \left(- s_i + n_i \left( 1 + \frac{s_i}{b_i} \right) \right)
\end{eqnarray}
We rule out model parameters which give $\mathcal{Q} > 5.99$, which corresponds to the 95\% upper limit for two degrees of freedom (three signal regions, total cross section is one constraint).  The benchmark signal was generated using Pythia~8\cite{Sjostrand:2014zea} followed by weighting with efficiencies provided in the auxiliary material from \cite{Khachatryan:2014mea}.  For generating signal samples for our models, we implement our model in FeynRules\cite{Alloul:2013bka} and generate events in Madgraph5\cite{Alwall:2011uj} using the UFO interface\cite{Degrande:2011ua} followed by showering and hadronisation using Pythia~8.

\end{document}